\documentclass[twocolumn,resetfootnote]{aastex701}
\usepackage{amsmath}
\usepackage{gensymb}

\begin{document}

\title{A Delayed Rejection Reversible Jump Markov Chain Monte Carlo Method for Multi-Resolution Maps of the Stochastic Gravitational Wave Background with Pulsar Timing Array Data}

\author[0009-0003-3700-446X]{Taha T. Moursy}
\affiliation{Department of Physics and Astronomy, Texas Tech University, Box 41051, Lubbock, TX 79409, USA}
\email{}
\author[0000-0002-8826-1285]{Nihan S. Pol}
\affiliation{Department of Physics and Astronomy, Texas Tech University, Box 41051, Lubbock, TX 79409, USA}
\email{}

\begin{abstract}
Pulsar timing array anisotropy analyses often use a naive counting argument to set resolutions of inferred maps of the stochastic gravitational wave background (GWB). We present a data-driven method in the form of a delayed rejection reversible jump sampler tailored to multi-resolution pixel decompositions of the angular power density of the GWB. We also consider a rapid, frequentist alternative based on information criteria statistics. We verify our methods with a series of injection-and-recovery simulations, finding that the standard counting argument would lead to drastic overfitting of the data and that our data-driven methods reduce the number of parameters in the models by one to three orders of magnitude yet can achieve higher resolution than the standard counting argument when justified by the data. We make our sampler and frequentist method implementation available on GitHub.
\end{abstract}


\section{Introduction} \label{introduction}
Pulsar timing arrays \citep[PTAs,][]{Sazhin78, Detweiler79} are sensitive to low-frequency gravitational waves (GWs) produced by supermassive black hole binaries (SMBHBs), among other potential sources. PTA experiments are conducted through precision timing of millisecond pulsars. The observed times of arrival (TOAs) of electromagnetic pulses are compared to predicted TOAs generated through careful modeling of various physical phenomena and noise processes (e.g., \citet{NG15timing, NG15noise}), and the differences are analyzed primarily to search for GWs.

One of the GW analyses that can be performed with PTA data is inferring the angular power distribution of a stochastic gravitational wave background (GWB) (see, e.g., \citet{Romano17}). This is an important analysis because anisotropy in the GWB is decisive evidence for an SMBHB origin as opposed to other hypotheses such as GWs from inflation (see e.g., \citet{NG15np}), which predict statistically isotropic GWBs \citep{Caprini18}. Many anisotropy searches have been performed by PTA collaborations but none have reported evidence for anisotropy \citep{EPTAanis,NG15anis,MPTAanis,PPTAanis}.

Inferring the angular power distribution of the GWB necessitates choosing a cutoff in the expansion of basis functions. This cutoff is typically chosen using a counting argument that the number of model parameters should not exceed the number of data points. Specifically, frequentist anisotropy analyses use cross-correlation estimators between pairs of pulsars as the data, so the counting argument in the pixel basis typically takes the form of $N_\text{pix} \lesssim N_\text{cc}$, where $N_\text{pix}$ is the number of pixels in a HEALPix\footnote{http://healpix.sf.net} \citep{HEALPix} tessellation of the sky and $N_\text{cc}$ is the number of cross-correlations. In practice, analyses are typically performed by setting the number of model parameters as close to the number of cross-correlations as possible (within the constraints of the nature of the basis), i.e., $N_\text{pix}\approx N_\text{cc}$.

This counting argument is flawed for several reasons. One is that the cross-correlations are covariant and hence not ``independent measurements'' \citep{Romano21,Allen23,pfos}. Other issues are that it does not include any information about signals present in the data, noise in the PTA, or the geometry of the array. The geometry of the PTA, set by the sky positions of the pulsars in the array, defines the directional sensitivity to anisotropy for that PTA, and a non-uniform distribution of pulsars on the sky implies a similarly non-uniform sensitivity to anisotropy on the sky \citep{Oliver26, Moursy26}.

This non-uniform sky-sensitivity of PTAs has led to a new method by \citet{Grunthal26} in which one sets the cutoff resolution higher than the typical choice but then uses truncated singular value decomposition (SVD) to discard modes of the reconstruction which the PTA is not sensitive to. This aims to produce sky maps adapted to the local resolution of a PTA at each sky location. \citet{Agarwal26} use a similar method to address biases in the inferred angular power distribution due to using too few terms in the expansion \citep{Semenzato25}. While these methods improve on the standard counting argument, they introduce another manually implemented hyper-parameter since one must select a threshold for the SVD truncation.

Regardless of the method, all previous works on the subject of inferring sky maps from PTA data use several arbitrary choices, including the maximum resolution and any regularization thresholds. In this work, we propose an entirely data-driven method using the reversible jump Markov chain Monte Carlo (RJMCMC; \citet{Green95}) algorithm and multi-resolution HEALPix meshes to simultaneously infer resolutions and angular power densities at each sky location, automatically incorporating in our inference the geometry and noise properties of the PTA as well as any signals and covariances that may be present in the data.

This paper is organized as follows. In Section \ref{sec:methods}, we describe relevant aspects of PTA data analysis, multi-resolution maps and reversible jump methods. We also describe a rapid but less robust and less informative method based on information criteria statistics. We end the section by describing the simulations we use to verify our methods, and we present in Section \ref{sec:results} our results of analyzing these simulations with our methods. We discuss our results and conclude with a brief summary in Section \ref{sec:discussion}.

\section{Methods} \label{sec:methods}
\subsection{Frequentist Map-Making with PTA Data}\label{sec:frequentist_maps}
We first give a summary of the commonly used cross-correlation method for inferring an angular distribution of the GWB using PTA data. This method has been used in PTA collaboration searches for anisotropy in the GWB \citep{NG15anis, MPTAanis, PPTAanis}. We will use this method for this work, but we note that the sampler we develop does not depend fundamentally on this method and could be modified to handle different methods of constructing GWB maps, such as using the TOAs directly \citep{Taylor20}.

Letting the angular power density of the GWB in direction $\hat\Omega$ be $\mathcal{P}(\hat\Omega)$, the GWB induces spatial correlations in the pulsar timing residuals, which are the differences between measured and expected times-of-arrivals (TOAs) of the emission from the pulsar. The expectation value of the correlations is the model-specific overlap reduction function (ORF). The ORF corresponding to a purely isotropic GWB is the well-known Hellings \& Downs (HD) curve \citep{HD}. The general formula of the ORF $\Gamma_{ab}$ for two pulsars $a$ and $b$, given $\mathcal{P}(\hat\Omega)$ and the detector response $\mathcal{R}_{ab}(\hat\Omega)$, is the convolution over the sky of $\mathcal{P}$ with $\mathcal{R}_{ab}$ \citep{orf_citation}
\begin{equation}
\Gamma_{ab} = \int\mathcal{R}_{ab}(\hat\Omega)\mathcal{P}(\hat\Omega)d\Omega.
\end{equation}

One can thus restrict the analysis to using just the inter-pulsar correlations rather than the full TOA basis with the advantage that the former is a significantly smaller dataset and allows lower computational turnaround time when searching for anisotropy \citep{Pol22}.
This model is often applied in anisotropy analysis pipelines with the following log-likelihood
\begin{equation}
    \log \ \mathcal{L}(\rho|\Gamma) = -\frac{1}{2}(\rho-\Gamma)^TC^{-1}(\rho-\Gamma)
    -\frac{1}{2}\log |C|
    -\frac{N_\text{cc}}{2}{\log2\pi}
    \label{eq:likelihood}
\end{equation}
where $\rho$ is the vector of measured pulsar pair cross-correlations, $C$ is the covariance matrix of the cross-correlations and $N_\text{cc} = N_\text{psr}(N_\text{psr}-1)/2$ is the number of cross-correlations given $N_{\rm psr}$ pulsars in the PTA.

In this pipeline, the cross-correlations are computed using the PTA optimal statistic \citep[OS,][]{Anholm09} framework. The OS is a statistic used to estimate the inter-pulsar cross-correlations, which can then be used to estimate the amplitude of the GWB in PTA data using the HD ORF as a template to fit the estimated cross-correlations. This framework was extended by \citet{pfos} to calculate frequency-resolved cross-correlations. This statistic is called the per-frequency OS (PFOS) and is advantageous for anisotropy searches as it enables the production of narrowband GWB sky-maps \citep{Gersbach25}. Narrowband anisotropy is a more realistic expectation if the source of the anisotropy is a loud SMBHB emitting approximately monochromatic GWs. 

In addition to using the PFOS in our work, we also account for the covariance of the cross-correlations induced by the presence of a GWB \citep{Romano21,Allen23,Johnson24,pfos}. Modeling the covariance of the cross-correlations improves the accuracy of GWB sky-map reconstruction \citep{Gersbach25}. The cross-correlations and covariance matrices are computed as given by Equations 28 and C21-C23 of \citet{pfos}, and we use the software package \texttt{Defiant} \citep{pfos} for these inputs to our likelihood in Equation \ref{eq:likelihood}.

\subsection{Multi-Resolution Maps}
In this work, we use the multi-resolution pixel basis introduced in \citet{Moursy26} with the goal of determining a preferred HEALPix tessellation of the sky given a PTA dataset. The multi-resolution pixel basis decomposes the GWB angular distribution $\mathcal{P}(\hat{\Omega})$ into a set of HEALPix pixels. For a decomposition with e.g. 12 pixels, there are 12 parameters in the model, and the estimated angular distribution vector is simply a 12-element vector with each element being an estimate of the GWB density at the center of the corresponding pixel ($\mathcal{P}(\hat\Omega_k)$ for pixel $k$). 

The multi-resolution pixel basis makes use of the multi-order-coverage \citep[MOC,][]{Fernique14, Reinecke15, Youngren17} method developed by \citet{Singer16} with applications to LIGO \citep{LIGO} data analysis. These methods were further developed by \citet{mhealpyPaper}) and implemented as a Python software package, \texttt{mhealpy} \citep{mhealpySoftware}, which provides convenient routines for manipulating and plotting these maps.

The multi-resolution pixel basis for PTA mapping of the GWB makes use of the NESTED and UNIQ HEALPix schemes for pixel identification, with the UNIQ scheme being the driving technology behind MOCs. The NESTED scheme works by identifying all pixels of a given resolution with the identifiers 0 through $N_\text{pix}-1$, where $N_\text{pix}$ is the number of pixels. A desirable quality of the NESTED scheme is that, for a pixel with a given NESTED index, $i_\text{NESTED}$, the four sub-pixels at the next resolution are given by $4i_\text{NESTED}, 4i_\text{NESTED}+1, 4i_\text{NESTED}+2$, and $4i_\text{NESTED}+3$. These sub-pixels are the pixels which together cover the area originally covered by the pixel at the previous resolution with index $i_\text{NESTED}$. The NESTED scheme thus makes querying pixels of different resolution trivial through simple multiplication and division by 4. 

The NESTED scheme uses the indices 0 through $N_\text{pix}$ for every given resolution. Therefore, knowing a pixel has index e.g. 0 does not uniquely identify the pixel. One also needs to know the resolution, either through the total number of pixels or some resolution parameter. The UNIQ scheme solves this by uniquely mapping the integers (greater than 3) to every possible pixel at all resolutions with a single mapping. It does this by not resetting the index to 0 after stepping to another resolution. One caveat is that it begins the indices from 4. So, the first 12 pixels at the lowest resolution of the HEALPix tessellations are numbered 4-15. The next 48 pixels (which are at the next resolution) are numbered 16-63, and so on. The relation between the UNIQ and NESTED indices of a pixel is
\begin{equation}
i_\text{UNIQ} = i_\text{NESTED}+4N_\text{side}^2,
\end{equation}
where $N_\text{side}$ is a resolution parameter. $N_\text{side}$ specifically is equal to the number of sub-pixels along one edge of a pixel at the lowest HEALPix resolution. For example, splitting a lowest-resolution pixel once gives four sub-pixels, with two along each edge of the original pixel, and so these sub-pixels have $N_\text{side}=2$. Possible values of $N_\text{side}$ in the UNIQ scheme are powers of 2, beginning with $2^0$. 

Coming back to PTA data analysis, we can choose any valid set of pixels to constitute our multi-resolution pixel basis, compute the detector response matrix $\mathbf{R}$ in the basis, and evaluate the log-likelihood given in Equation \ref{eq:likelihood}. The detector response matrix has elements
\begin{equation}
    R_{ab}^k=\frac{\Delta\Omega_k}{4\pi}\left[\mathcal{F}^\times_a(\hat\Omega_k)\mathcal{F}_b^\times(\hat\Omega_k)+\mathcal{F}_a^+(\hat\Omega_k)\mathcal{F}_b^+(\hat\Omega_k)\right],
\end{equation}
where $\Delta\Omega_k$ is the area of pixel $k$ and $\mathcal{F}^A_a(\hat\Omega_k)$ is the response of pulsar $a$ to a GW of polarization $A$ evaluated at the location $\hat\Omega_k$, which is the center of pixel $k$ (see, e.g., \citet{NG15anis}).
Once one selects a multi-resolution tessellation, one can simply maximize the likelihood shown in Equation \ref{eq:likelihood} for a frequentist estimate of the GWB angular power distribution.
However, it is not obvious what resolution GWB sky-maps should be reconstructed with using PTA data, so we use an RJMCMC method to simultaneously infer the preferred number, resolution, and amplitudes of pixels in a GWB sky-map.

\subsection{Reversible Jump MCMC}\label{sec:rjmcmc}
The RJMCMC algorithm, proposed by \citet{Green95}, is a reframing of Metropolis-Hastings \citep[MH,][]{Metropolis53, Hastings70} which enables an MCMC sampler to move across models having different parameter spaces and, in general, different dimensionality.
The RJMCMC algorithm works as follows.
Let $x$ denote the current state, represented by a vector of parameter values for the current model and some label specifying the model. Let $y$ denote a proposed state for a new model and associated parameter values. The RJMCMC algorithm then consists of drawing a vector $u$ of auxiliary variables from a density $q$, mapping ($x$, $u$) through a diffeomorphism $T$ to the proposed state $y$, and accepting it with probability \citep{Green95}
\begin{equation}
    \alpha(x,y) = \text{min}\left\{ 1, \frac{\pi(y)}{\pi(x)}\frac{q'(u')}{q(u)}\frac{j'(x|y)}{j(y|x)}\left|\frac{\partial(y,u')}{\partial(x,u)}\right| \right\},
    \label{eq:RJ_acceptance_ratio}
\end{equation}
where $u'$ is the vector of auxiliary variables drawn from the density $q'$ for the reverse move $y \rightarrow x$ such that ($y$, $u'$) is mapped to ($x$, $u$) by the inverse diffeomorphism $T^{-1}$. $u$ and $u'$ also serve to satisfy the dimension-matching condition of the RJMCMC algorithm, which states the number of dimensions in the model of state $x$ plus the number of auxiliary variables drawn in $u$ must equal the corresponding sum of dimensions in the model of state $y$ and the number of auxiliary variables in the vector $u'$. This condition is necessary for $T$ to be invertible or, equivalently, bijective. $\pi$ includes priors on the models themselves, priors on the parameters of the models, and the likelihood. $j$ and $j'$ are the model proposal kernels from which $y$ and $x$ are drawn, respectively. Finally, the absolute value of the determinant of the Jacobian accounts for the change of variables.

In our application of multi-resolution HEALPix maps, we use a reversible pair of proposals that consists of (1) merging four pixels of a given $N_\text{side}$ into one pixel of resolution $N_\text{side}/2$ and (2) splitting a single pixel of resolution $N_\text{side}$ into four sub-pixels with resolution $2N_\text{side}$.

As a concrete example, we show explicitly how a split-move is performed in our application. In determining the acceptance ratio, we will also need to consider at the same time the hypothetical merge-move which reverses our split-move.
Given a state $x$, uniquely identified by the collection of amplitudes and $\text{UNIQ}$ indices of its constituent pixels, we propose a new state $y$ which is identical to $x$ except one of the pixels in $x$ has been removed and replaced with the four pixels of the next resolution. Three probabilities go into proposing $y$ from $x$: (1) choosing to propose a split-move, (2) choosing which pixel to split, and (3) determining the amplitudes of the new pixels, which involves the auxiliary variables $u$. Our $j$ probability is the product of (1) and (2). We set the probability of proposing a split-move equal to the merge-move proposal probability, so (1) contributes only a factor of 1 to $j'(x|y)/j(y|x)$. For (2), we set the probability of choosing the particular pixel to split to be 1 out of the total number of available pixels that can be split (i.e., equal weights), which is often the total number of pixels in $x$ except we set a model prior in the form of a resolution cap $N_\text{side} < 128$. So far, we have $j(y|x)=1/N_{\text{pix},x}$, assuming no pixels are at the resolution cap, which is done to simplify this example but not assumed in the corresponding code in our sampler. For (3), we draw four auxiliary variables comprising the vector $u$ independently from $q$, so the probability of drawing $u$ is the product of $q$ evaluated at each auxiliary variable. We describe our choice for $q$ in Section \ref{sec:sampler}.
After determining a proposed state $y$, we evaluate the prior (on both the proposed model and the proposed parameters) and the likelihood to compute $\pi(y)$.

One must then determine the probabilities for reversing the move in order to compute the acceptance ratio. We determine the auxiliary variables $u'$ which would have to be drawn to propose $x$ from $y$ and evaluate the probability as $q'(u')$. For the probability of proposing the model of state $x$ from $y$ (i.e., $j'(x|y)$), we use our merge-move transition kernel, which works by randomly selecting a pixel from the set of available pixels. The set of available pixels is the intersection of the set of pixels with resolution greater than $N_\text{side}=1$ and the set of pixels with all their sister pixels in the current model. To clarify this second set, compare the two sets of pixels $\{16,17,18,19\}$ and $\{64,65,66,67,17,18,19\}$: in the first set, all four pixels can be selected for a merge-move, while only the pixels $\{64,65,66,67\}$ would be valid choices from the second for a merge-move. This is required because our split-move transition kernel only generates single increments in resolution, so attempting to merge the seven pixels in $\{64,65,66,67,17,18,19\}$ would give $j'(x|y)=0$. We emphasize this is not fundamental to RJMCMC but is due to our implementation as we could simply choose a more complex transition kernel which allows for larger jumps in resolution.

To summarize, we have $\pi(y)$, $\pi(x)$, $q'(u')/q(u)$, and $j'(x|y)/j(y|x)$, although we have not discussed our choice of $q$ and $q'$ yet as we will do so in Section \ref{sec:sampler}. This leaves the Jacobian in the acceptance ratio, which we explain next.

The diffeomorphism we use, $T$, maps the pixels which are the same between states $x$ and $y$ to each other (i.e., their amplitudes and resolutions remain the same after a merge or split move) and maps the single auxiliary variable $u'$ to the pixel to be split and the auxiliary vector $u$ to the four new pixels. In this way, the Jacobian of $T$ is the identity matrix for the first $N_{\text{pix},x}-1$ pixels, but the last five rows and columns of the Jacobian are four row-swaps from the identity. Therefore, the determinant of this Jacobian is $(-1)^4=1$. Explicitly,

\begin{equation}
    J = 
    \begin{pmatrix}
        I_{N_\text{pix-1}} & 0 \\
        0 & \mathcal{J}
    \end{pmatrix},
\end{equation}

where

\begin{equation}
\mathcal{J}=
\begin{pmatrix}
\frac{\partial{\theta'_{N_\text{pix}}}}{\partial{\theta_{N_\text{pix}}}} & \frac{\partial{\theta'_{N_\text{pix}}}}{\partial{u_1}} & \frac{\partial{\theta'_{N_\text{pix}}}}{\partial{u_2}} & \frac{\partial{\theta'_{N_\text{pix}}}}{\partial{u_3}} & \frac{\partial{\theta'_{N_\text{pix}}}}{\partial{u_4}} \\
\frac{\partial{\theta'_{N_\text{pix}+1}}}{\partial{\theta_{N_\text{pix}}}} & \frac{\partial{\theta'_{N_\text{pix}+1}}}{\partial{u_1}} & \frac{\partial{\theta'_{N_\text{pix}+1}}}{\partial{u_2}} & \frac{\partial{\theta'_{N_\text{pix}+1}}}{\partial{u_3}} & \frac{\partial{\theta'_{N_\text{pix}+1}}}{\partial{u_4}}\\
\frac{\partial{\theta'_{N_\text{pix}+2}}}{\partial{\theta_{N_\text{pix}}}} & \frac{\partial{\theta'_{N_\text{pix}+2}}}{\partial{u_1}} & \frac{\partial{\theta'_{N_\text{pix}+2}}}{\partial{u_2}} & \frac{\partial{\theta'_{N_\text{pix}+2}}}{\partial{u_3}} & \frac{\partial{\theta'_{N_\text{pix}+2}}}{\partial{u_4}}\\
\frac{\partial{\theta'_{N_\text{pix}+3}}}{\partial{\theta_{N_\text{pix}}}} & \frac{\partial{\theta'_{N_\text{pix}+3}}}{\partial{u_1}} & \frac{\partial{\theta'_{N_\text{pix}+3}}}{\partial{u_2}} & \frac{\partial{\theta'_{N_\text{pix}+3}}}{\partial{u_3}} & \frac{\partial{\theta'_{N_\text{pix}+3}}}{\partial{u_4}}\\
\frac{\partial{u'_1}}{\partial{\theta_{N_\text{pix}}}} & \frac{\partial{u'_1}}{\partial{u_1}} & \frac{\partial{u'_1}}{\partial{u_2}} & \frac{\partial{u'_1}}{\partial{u_3}} & \frac{\partial{u'_1}}{\partial{u_4}}
\end{pmatrix} .
\end{equation}

Setting $u = (\theta'_{N_\text{pix}},...,\theta'_{N_\text{pix}+3})$ and $u'= \theta_{N_\text{pix}}$ and then evaluating the derivatives gives
\begin{equation}
\mathcal{J}=
    \begin{pmatrix}
    0 & 1 & 0 & 0 & 0 \\
    0 & 0 & 1 & 0 & 0 \\
    0 & 0 & 0 & 1 & 0 \\
    0 & 0 & 0 & 0 & 1 \\
    1 & 0 & 0 & 0 & 0
    \end{pmatrix} .
\end{equation}

This concludes our description of the split-move, and the merge-move is simply the inverse. We additionally include a second pair of split and merge moves in which the entire map is updated rather than keeping the common pixels between models at the same amplitude. The proposed values are drawn independently of the current state of the sampler, so in this case we take $u=\theta'$ and $u'=\theta$, and so the Jacobian is the identity matrix.

\subsection{Delayed Rejection}\label{sec:delayed_rejection}
The standard MH algorithm accepts a proposed move with the MH acceptance ratio. If the proposed move is rejected, the current state is repeated in the chain and the algorithm continues. However, this algorithm can lead to inefficiency by increasing the auto-correlation length of the MCMC. Delayed rejection is a method that aims to improve the efficiency of MCMCs by allowing for more proposed moves (which may depend on the rejected move) after a move is rejected. 

The delayed rejection method was proposed by \citet{Tierney99} for MH and adapted for RJMCMC by \citet{Green01}. The acceptance ratio for the RJMCMC algorithm, Equation \ref{eq:RJ_acceptance_ratio}, when generalized to include delayed rejection and applied to second-stage delayed rejection, is \citep{Green01}
\begin{multline}
    \alpha_2(z,x) = \text{min}\Biggl\{ 1,\frac{\pi(z)}{\pi(x)} \frac{g_1'(u_1',x|y^*) g_2'(u_2',y^*|z)}{g_1(u_1,y^*|x)g_2(u_2,z|y^*)} \\ \times \frac{1-\alpha_1(z,y^*)}{1-\alpha_1(x,y)} \left|\frac{\partial(z, u_1', u_2')}{\partial(x, u_1, u_2)}\right|\Biggr\},
\end{multline}
where $z$ is the proposed state at the second stage (i.e., after rejecting model $y$), $u_1'$ and $u_2'$ are auxiliary variables drawn in going from $z$ to $x$, $y^*$ is an intermediate state between $x$ and $z$ that may be different from $y$, and $\alpha_1(z,y^*)$ is given by Equation \ref{eq:RJ_acceptance_ratio} with the appropriate substitution, replacing $x$ and $y$ in that equation with $z$ and $y^*$, respectively. We use $g(u,y|x)$ following \citet{Green01} as an abbreviation for $q(u)j(y|x)$.

In what follows, we restrict ourselves to the case that $y=y*$.
The specific details of applying this equation to our application are essentially identical to those given under Equation \ref{eq:RJ_acceptance_ratio} but with appropriate replacements of variable names, e.g., using $z$ instead of $x$ when computing $\alpha_1(z,y)$.

In our application, we use delayed rejection in the case that an inter-model proposal is rejected. We propose a new model that is the result of applying the rejected proposal type to the rejected proposed model. For example, if a split-move is rejected, we apply a second split to the rejected mesh and propose the resulting model, and we do likewise for merge-moves. This enables the sampler to move two steps in resolution in a single iteration, as well as enabling move types where different pixels around the map are merged or split.

Note that, while the delayed rejection algorithm can be nested to generate an arbitrary number of proposed moves, we find that our sampler performs well with only one iteration of delayed rejection. In addition, as discussed by \citet{Green01}, nested iterations can be computationally expensive, so the trade-off between auto-correlation length and computational runtime should be considered when designing a delayed rejection sampler.

\subsection{Sampler}\label{sec:sampler}
We implement the above methods in \texttt{Resolve}, an open-source Python package accompanying this work and available on PyPI and GitHub\footnote{github.com/TTMoursy/Resolve}. The sampler is accelerated with \texttt{JAX} and \texttt{Numba} and makes use of the Levenberg-Marquardt \citep[LM,][]{Levenberg44,Marquardt63} algorithm as implemented in \texttt{JAXopt}. We use this local optimization algorithm to obtain maximum-likelihood estimates (MLEs) which are used in evaluating Hessians and in generating random proposals for new pixel amplitudes, which we find to be much more efficient than drawing from the prior or using an approximate analytical solution. We cache the optimal solution to improve the speed and efficiency of our sampler so that the optimization problem is solved only once per model explored by the sampler.

For intra-model moves, simple symmetric draws from independent normal distributions for pixel angular power densities can be inefficient for some tessellations due to strong covariances between pixels, so instead we compute the Hessian of the log-likelihood with \texttt{JAX}'s automatic differentiation and draw proposals from a multivariate normal with a covariance equal to the inverse of the negative Hessian evaluated at the maximum-likelihood pixel amplitudes. However, we modify the Hessian before inverting because some pixels have upper-limit type posteriors and can therefore cause numerical instabilities when inverting the Hessian. We cache the Cholesky (or, if needed, singular-value) decomposition of the covariance matrix the first time this move is proposed in a model, allowing for rapid future proposals within the model.

We described much of our sampler's inter-model proposal mechanisms in Sections \ref{sec:rjmcmc} and \ref{sec:delayed_rejection}, but here we detail some choices we make that are not necessarily specific to RJMCMC. In particular, we discuss our auxiliary variable densities $q$ and $q'$ and our default priors, although the priors can be set by the user of our sampler.
When proposing a new model, we first perform numerical optimization of the likelihood using the LM algorithm as mentioned previously. We use numerical optimization because modeling the pixel amplitudes in log-space comes at the expense of no longer having an analytical maximum likelihood solution but has the benefits of ensuring positivity and improving scaling of the parameter space. To set an initial state for the optimizer, we use the standard, linear pixel basis and the Tikhonov-regularized analytical solution $P=(M+\lambda I)^{-1}X$, where $M=R^TC^{-1}R$ and $X=R^TC^{-1}\rho$ are the Fisher information matrix and dirty map, respectively, \citep{Romano17,AliHaimoud20,AliHaimoud21}, and we arbitrarily select $\lambda=10^{-3}$, which we find to balance well between stabilizing and biasing the initial estimate, helping the optimizer to converge more efficiently and reliably.

After obtaining the MLE for the pixel amplitudes for the proposed model, we draw auxiliary variables $u$ and $u'$ from $q$ and $q'$, which we construct as follows. We use two different density forms for these distributions, depending on whether the posterior of the pixel amplitudes is well-constrained or is only an upper limit. 
We classify pixels as having upper-limit posteriors during sampling by evaluating the log-likelihood at the MLE but with one pixel at a time having amplitude set to the lower prior bound. If the log-likelihood changes by less than $\frac{1}{2}$, we classify the pixel as having an upper-limit posterior and draw from the mixture described; otherwise, we classify the pixel as having a well-constrained amplitude and draw from a multivariate normal as previously described.
For pixels which are well-constrained, we draw their amplitudes from a multivariate normal distribution centered on the MLE and with a covariance matrix obtained through a Fisher estimate and \texttt{JAX}-based automatic differentiation. 
For pixels with upper-limit type posteriors, we draw the amplitudes from a mixture of a uniform distribution from the lower prior bound up to an estimated upper limit for each pixel and a triangular distribution with vertices at the prior bounds and mode at the lower prior bound (i.e., a straight line with a negative slope from one end of the prior to the other). We estimate the upper limits for each upper-limit-type pixel by changing its value while holding the other pixel amplitudes fixed at the MLE and determining values where the log-likelihood decreases by $\frac{1}{2}$. For more precise estimates, one would use the profile likelihood to account for covariances between pixels, but this comes with the added computational cost of solving many optimization problems, and we are interested only in a rough estimate for the proposal density. We find this approximation works well in practice as can be seen in Section \ref{sec:results}. 

Our priors for pixel angular power densities are taken as independent and uniform in log-space over $[-7,7]$, and we adopt a uniform prior on models. We impose a maximum $N_\text{side}$ of 64, although this seems unlikely to have affected our results since the preferred models have low values of $N_\text{side}$. These prior choices can each be specified by the user before running our sampler in \texttt{Resolve}, but note that $N_\text{side}$ must be a power of two since our method is based on the UNIQ HEALPix scheme (which is based on the NESTED scheme). These schemes allow for trivial pixel querying at different resolutions by simply multiplying or dividing pixel indices by factors of four. This requires that $N_\text{side}$ only be incremented by factors of two. In contrast, the RING scheme (not used in this work) has neither the restriction on $N_\text{side}$ nor the convenient querying routines for changing resolution.

\subsection{Post-Processing}
After obtaining a chain from an RJMCMC simulation (with burn-in discarded), one can estimate the posterior odds ratio between two models by taking the ratio of the number of samples that correspond to each of the two models. If one uses uniform priors on the models, this estimate also serves as an estimate for the Bayes factor between the two models. In the general case, one can obtain an estimate for the Bayes factor by dividing the posterior odds ratio estimate by the prior model ratio; this allows for convenient model selection independent of the model priors. We use this method to compute all of the Bayes factors reported in this work, and we use bootstrap resampling to estimate uncertainties on the Bayes factors.

\subsection{A Rapid Frequentist Method}\label{sec:bic}
Performing an RJMCMC analysis with our sampler can take from a few hours to several days, depending on the complexity of the posterior and the desired number of samples, so we also present a rapid but purely frequentist alternative based on the Akaike information criterion (AIC), $2N_\text{pix}-2\log\mathcal{L}$, and Bayesian information criterion (BIC), $N_\text{pix}\log N_\text{cc}-2\log\mathcal{L}$. This method we propose is based on the algorithm presented by \citet{Singer16} but is simplified and works by iteratively splitting some threshold percentage of the integrated angular power density MLE, computing the AIC or BIC for each mesh, and selecting as the preferred model the one which minimizes the AIC or BIC. This alternative method does not provide Bayes factors between models and introduces some arbitrary choices (and thereby losing the purely data-driven aspect of RJMCMC), but it is much faster, taking only a few seconds for an analysis or a few minutes if one also computes uncertainties on the angular power density. 

We estimate uncertainties on the parameters using a profile likelihood method, whereby we use a root-finding algorithm to find the density which decreases the profile likelihood by $\frac{1}{2}$ from the maximum likelihood. At each iteration of the root-finding algorithm, we perform optimization of all other pixels using the limited-memory Broyden-Fletcher-Goldfarb-Shanno (LBFGS) algorithm \citep{Nocedal06, Sorber12}; therefore, these uncertainties account for the covariance between pixels and is more robust than the approximate likelihood scan we describe in Section \ref{sec:sampler} although this is more computationally expensive. The root-finding algorithm we use is bisection, and both the bisection and LBFGS algorithms we use are implemented in \texttt{JAXopt}, which comes with the benefit of being compatible with automatic parallelism through \texttt{JAX} so that batches of independent optimization problems can be solved simultaneously.

\subsection{Simulations}\label{sec:simulations}
To verify our method, we create two sets of simulations, one idealistic and the other realistic. The idealistic set consists of two simulations which have data simulated at the cross-correlation level. The simulated PTA for these datasets consists of 75 pulsars distributed randomly across the sky; the sky positions can be seen in Figure \ref{fig:cc_results}. The first simulation in this set is generated by drawing cross-correlations from independent normal distributions centered at the HD coefficients and with standard deviation 0.1 (the maximum HD cross-correlation has value 0.5, for reference). This simulation is meant to represent an isotropic GWB with a small amount of noise in the measurements. For the second idealistic simulation, we would like to test the ability of the sampler in recovering multiple hotspots and well-localized anisotropies. So, we generate this simulation by choosing a mesh and pixel amplitudes and calculating the theoretical cross-correlations such a GWB would produce. We use cross-correlation uncertainties of 0.1, but we do not add any noise to the theoretical cross-correlations for this second simulation. We include two pixels with angular power densities greater than 1, and we set the resolution of one of these pixels to $N_\text{side}=4$ and the other to $N_\text{side}=16$. The injected mesh is shown in the bottom left panel of Figure \ref{fig:cc_results}. Note that the injected angular power density is $100$ for the $N_\text{side}=4$ pixel and $1600$ for the $N_\text{side}=16$ pixel, which are equivalent pixel amplitudes and correspond to an angular power density of $\sim6$ at $N_\text{side}=1$.

The second set of simulated datasets is more realistic and generated at the TOA-level. The simulations are based on the NANOGrav 12.5-yr dataset \citep{NG12p5}. We generate simulated versions of the 45 pulsars used in the NANOGrav GWB search \citep{NG12p5gwb} of that dataset. Our simulated pulsars have the same timing models as the real dataset but include some simplifications. We simulate three datasets with various signals injected: a GWB in all three datasets and a CW with different chirp mass in two of the datasets. All of the CW parameters are identical in the two datasets except for the chirp mass; one CW has a chirp mass of $10^9 M_\odot$, and the other has a chirp mass of $10^{10} M_\odot$. The other parameters for the CWs are $f_\text{GW}=6.8/T_\text{span}\approx15$ nHz, $\theta=2.13$ rad, $\phi=3.57$ rad, $d_L=55$ Mpc, $\psi=\pi/3$ rad, and $\iota=\pi/4$ rad.
We inject measurement noise into the simulated datasets in the form of EFAC, EQUAD, and ECORR, using the same values measured in the real dataset. We also inject intrinsic pulsar red noise with a power-law (PL) power spectral density (PSD), again using the same values measured in the real dataset. The injected GWB is isotropic and has a PL PSD with spectral index $\gamma=13/3$ and amplitude $A=2.4\times10^{-15}$ at a reference frequency of $f=1/\text{yr}$, taken from the results of \citet{NG15gwb} and generally consistent with the range of results from the various PTA collaborations.

\section{Results}\label{sec:results}

\subsection{Idealistic Simulations}\label{sec:results_simple_simulations}
\begin{figure*}
    \includegraphics[width=\textwidth]{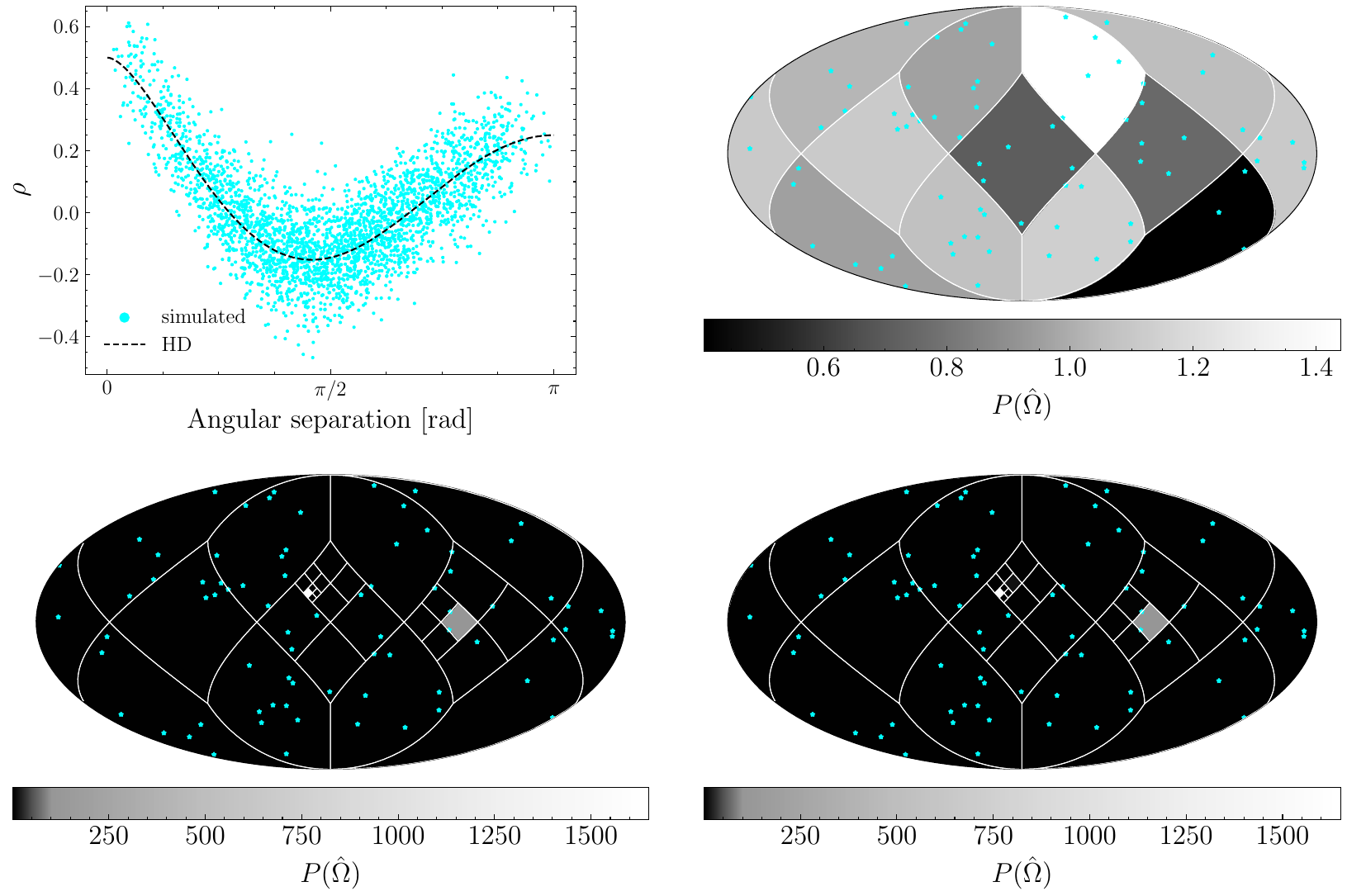}
    \caption{Injection and recovery summary plots for the idealistic datasets. The left column represents the injections, and the right column shows the preferred mesh from each analysis with the plotted angular power densities being point estimate summary statistics taken from the posteriors. The point estimate we use is the mode of the 1d-marginalized posterior of each pixel. \textit{Top Row: }The simulation in which the cross-correlations are drawn from independent normal distributions around the HD curve, representing an isotropic GWB with noise. \textit{Bottom Row: }The simulation in which the exact analytic cross-correlations are used in order to test the ability of the sampler in recovering a high-resolution anisotropy as well as multiple hotspots. This simulation involves a base isotropic GWB with two pixels of increased angular power density and at higher resolutions. The preferred mesh and inferred angular power densities shown in the right column indicate the sampler succeeded in both simulation tests. The simulation represented in the bottom row is the same as that represented in the left column of Figure \ref{fig:diagnostics}.
    The cyan markers in the Mollweide projections show the positions of the simulated pulsars.}
    \label{fig:cc_results}
\end{figure*}

\begin{figure*}
    \includegraphics[width=\textwidth]{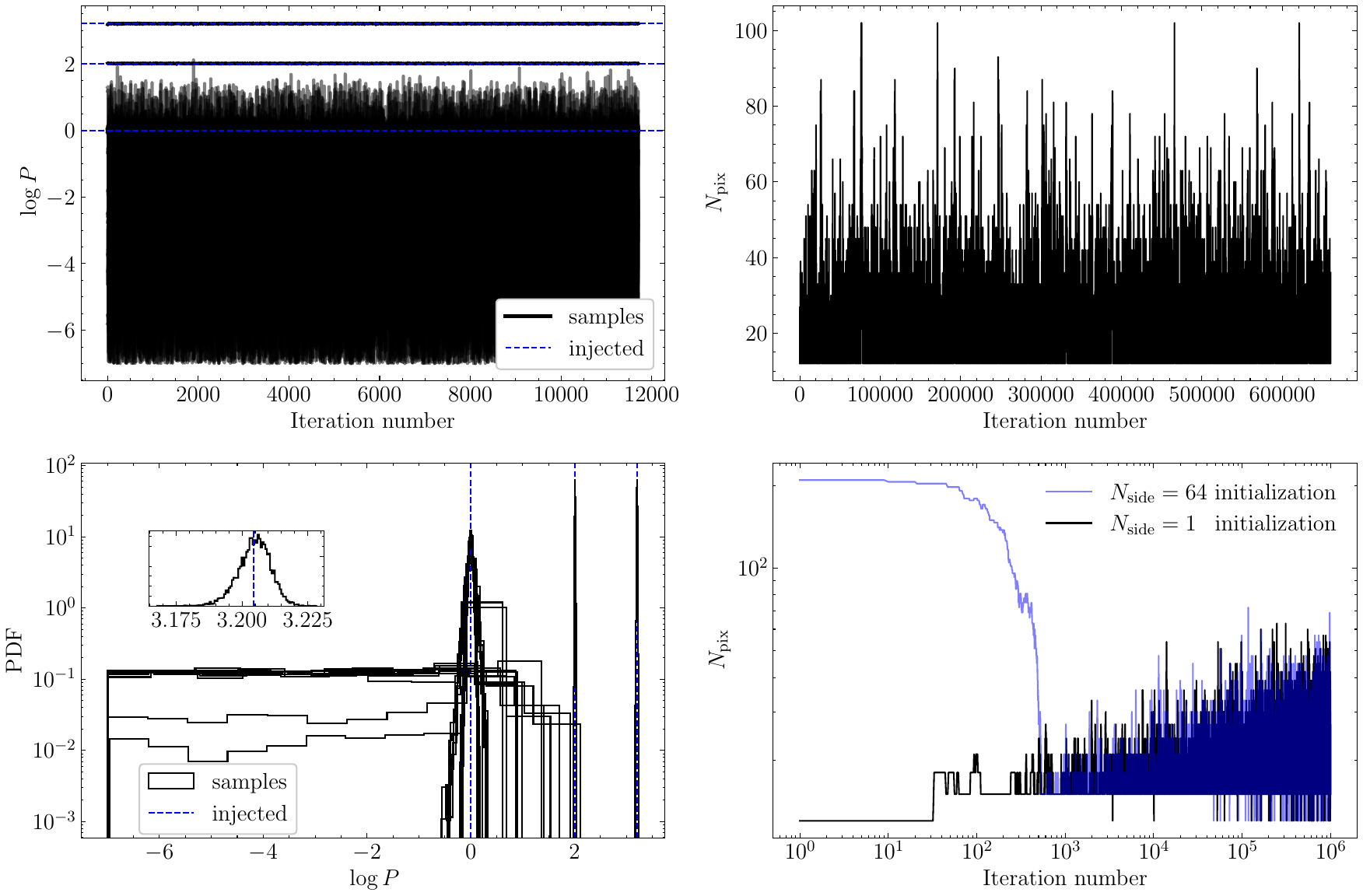}
    \caption{Diagnostics of the sampler's performance.
    The left column represents posteriors (after thinning) for the angular power density of the preferred mesh for one of the simulations used in this work.
    \emph{Top left:} Trace plots of the angular power density of each pixel in the mesh, with the recovered values shown in black and the injected values shown with dashed blue lines. This shows the sampler achieves good mixing within a given model.
    \emph{Bottom left:} Histograms of the 1d-marginalized posteriors for each pixel, and an inset plot showing the histogram for the pixel with the largest angular power density, better showing the shape of the posterior. The notable differences in shape of posterior indicate the importance of using well-designed transition kernels.
    \emph{Top right:} Number of pixels as a function of iteration number for one of the simulations in this work. The number of pixels changes rapidly through the sampling, indicating the sampler achieves good mixing between models. Note also that the number of pixels does not correspond one-to-one to the different models; in general, each value of $N_\text{pix}$ corresponds to many distinct meshes. For example, in the chain here, there are $\sim$47000 meshes visited by the sampler.
    \emph{Bottom right:} Number of pixels as a function of iteration number of the sampler for two different chains. One chain is initialized with $N_\text{side}=1$, and the other chain is initialized with pixels near the injected CW source having $N_\text{side}=64$. The two chains become indistinguishable within a few hundred iterations despite the much higher initial resolution. This shows that the sampler can rapidly and efficiently traverse inter-model space when initialized far from the preferred mesh.}
    \label{fig:diagnostics}
\end{figure*}

We show in Figure \ref{fig:cc_results} the injection and recovery results of using our method with the idealistic simulations. For the simulation which is meant to mimic an isotropic GWB by drawing the correlations from normal distributions centered on the HD curve, we find the preferred mesh is the lowest HEALPix resolution, with the amplitude posteriors all consistent with an isotropic GWB. For the simulation which uses the exact analytic cross-correlations produced by the anisotropic GWB represented in the lower left panel of the figure, the preferred mesh is identical to the injected mesh, and the recovered pixel amplitudes are consistent with the injections. Since the Mollweide projections require a point estimate of the pixel amplitudes rather than being able to represent the whole posterior, we show trace plots and histograms of 1d-marginalized posteriors of the pixel angular power densities in the left column of Figure \ref{fig:diagnostics} for comparison between the injected and recovered amplitudes. It can be seen from that column that the sampler achieves good mixing within the model and that the recovered amplitudes are consistent with the injections, while the right column of Figure \ref{fig:diagnostics} shows good mixing between models. This simulation verifies the ability of our sampler in recovering multiple pixels of excess angular power density and to recover appropriate resolutions for each region of excess angular power density (in this case, one at $N_\text{side}=4$ and one at $N_\text{side}=16$).

\subsection{Realistic Simulations}\label{sec:results_realistic_simulations}
To mimic the real data analysis pipeline used in PTA map-making of the GWB, we also generate simulations starting at the level of the TOAs as described in Section \ref{sec:simulations}. Following standard PTA methods, we use \texttt{ENTERPRISE} and \texttt{PTMCMCSampler} to conduct a broken power law (BPL) analysis and then a power law (PL) analysis for each simulation. We fix the measurement (``white'') noise parameters to their true values in both analyses. We use the BPL analysis to determine an appropriate number of Fourier modes for the GWB model, and we then estimate the red noise and GWB parameters from the PL analysis with the inferred number of modes. As in \citet{NG15anis, MPTAanis, Gersbach25}, we next use the pair-covariant PFOS framework to estimate cross-correlations and the covariance matrix as described in Section \ref{sec:frequentist_maps}.

\begin{figure*}
    \includegraphics[width=\textwidth]{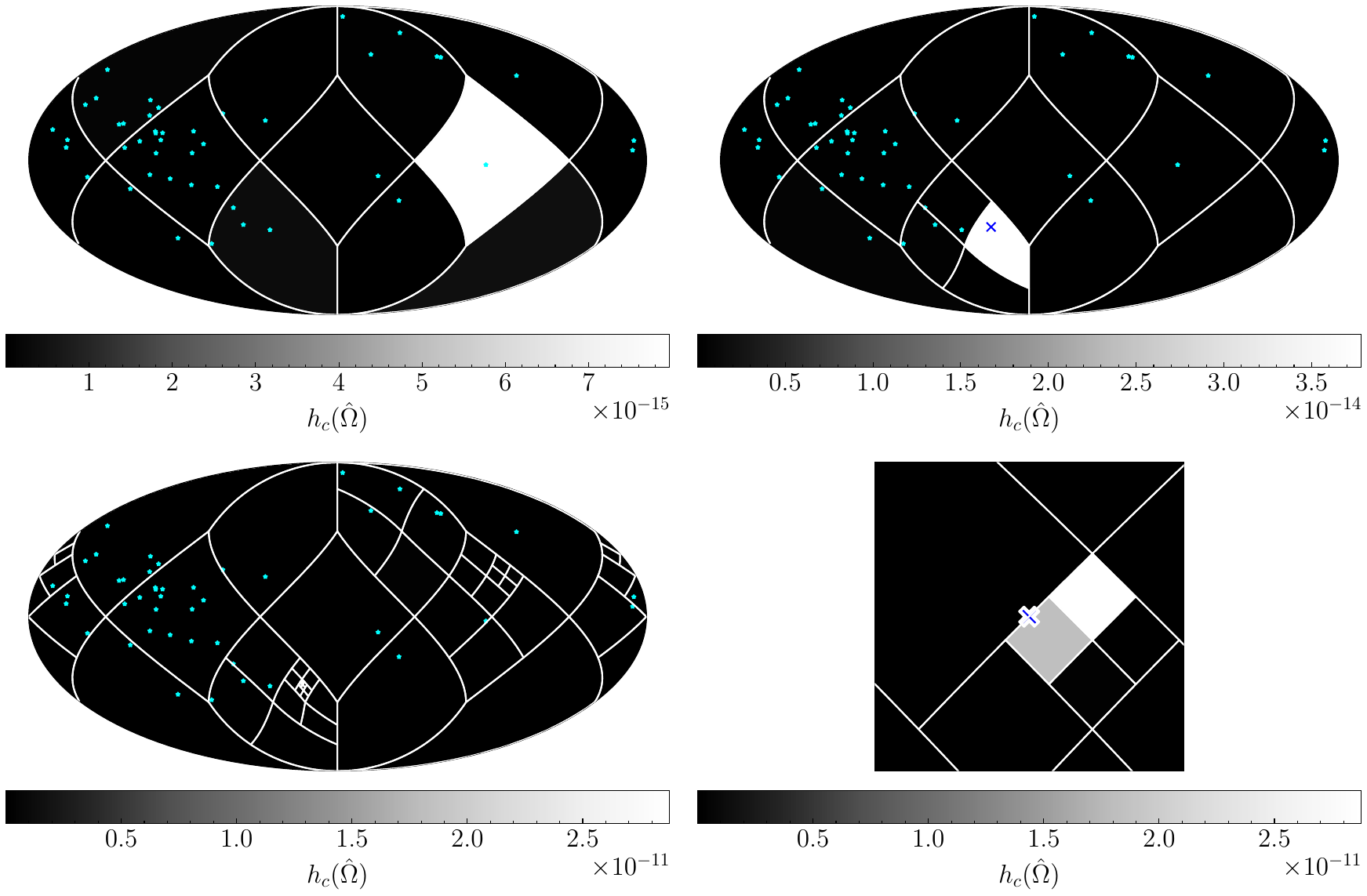}
    \caption{Projections of the preferred mesh from the realistic set of simulations. The top left plot corresponds to a simulation in which only an isotropic GWB was injected. The top right plot corresponds to a simulation in which a $10^9 M_\odot$ chirp mass binary was injected at the position marked with a blue x. The bottom plots correspond to a simulation in which a $10^{10} M_\odot$ chirp mass binary was injected (at the position marked with a blue x in the right plot), with the left plot being a Mollweide projection of the whole sky and the right plot being a $10^{\circ}\times10^{\circ}$ Cartesian projection centered on the location of the injected binary. Note that the quantity plotted here is the angular power density, so one must integrate these values over the sphere to get the characteristic strain of the GWB.
    Altogether, these results give further verification to the sampler and indicate that the counting argument conventionally used would lead to over-resolution in most sky locations and under-resolution when the signal is loud enough, emphasizing the importance of data-driven methods.}
    \label{fig:realistic_results}
\end{figure*}

Using the resulting cross-correlations and covariance matrix to compute the likelihood in Equation \ref{eq:likelihood}, we run our sampler for each of the three realistic simulations described in Section \ref{sec:simulations}. We show in Figure \ref{fig:realistic_results} the preferred models, with the pixel amplitudes shown being the modes of the 1d-marginalized posteriors. We find that for the isotropic GWB injection, the preferred mesh is the lowest resolution HEALPix mesh, and integrating the angular power density point estimate (shown in the Mollweide projection) over the sky gives a characteristic strain of $8\times10^{-15}$. Integrating each sample in the posterior gives a minimum characteristic strain of $2\times10^{-17}$ and maximum of $2\times10^{-14}$. The injected GWB has a characteristic strain of $4\times10^{-15}$ at this frequency.

For the simulation which includes an injected CW with chirp mass $10^9 M_\odot$, the preferred mesh is again the lowest resolution mesh except for the pixel containing the sky location of the injected CW, which is split once so that the anisotropy induced by the CW is localized to a pixel with resolution $N_\text{side}=2$. As an example of how our method produces Bayes factors, the preferred mesh has $\sim600,000$ samples in a chain of $10^6$ iterations, and the mesh which is the same but one step further in resolution (i.e., $N_\text{side}=4$ for the pixel containing the sky location of the CW) has $\sim4,000$ samples. The Bayes factor between these two meshes is $\sim150\pm2$ in favor of the mesh with the $N_\text{side}=2$ pixel, and the uncertainties represent $1-\sigma$ uncertainties from bootstrap resampling as mentioned in Section \ref{sec:sampler}. The base HEALPix mesh has even less posterior odds, with only $\sim800$ samples, and ranks as the 74th model; the Bayes factor between the preferred mesh and the base mesh is $720\pm30$. Integrating the point estimate shown for the angular power density shown in Figure \ref{fig:realistic_results} for this simulation gives a characteristic strain of $2\times10^{-14}$. The strain amplitude of the injected CW is $6\times10^{-15}$, although the frequency at which it is injected is at $6.8/T_\text{span}$ while the analysis results correspond to $7/T_\text{span}$. Integrating each sample of the posterior separately gives a minimum characteristic strain of $1\times10^{-14}$ and maximum of $3\times10^{-14}$.

The results for the $10^{10} M_\odot$ chirp mass injected binary simulation show a preferred mesh that achieves much higher resolution than the previous two simulations, with the anisotropy induced by the CW localized to a pixel of resolution $N_\text{side}=64$. Note that this $N_\text{side}$ is the upper bound of our prior, so we perform a second analysis in which we allow $N_\text{side}$ to reach $256$, but the preferred mesh is the same, so this result is not an artifact of the prior choice. Integrating the point estimate angular power density in Figure \ref{fig:realistic_results} gives a characteristic strain of $1\times10^{-12}$, and the injected strain amplitude of the CW is $3\times10^{-13}$. Integrating each sample in the posterior gives a minimum and maximum characteristic strain of $\pm1\%$ of the point estimate, so the injected strain amplitude is excluded from the posterior.

\subsection{Impact of Choice of Initial Mesh}
By default, our sampler begins with a single-resolution mesh with $N_\text{side}=1$, but the starting resolution or starting mesh can be specified by the user. Of course, it is less computationally expensive to begin a sampler near the mode of the posterior than in a region of low support. We would like to verify however that the performance of the sampler is independent of the starting mesh and that the inference is not biased by such an initialization. To do so, we reanalyze the realistic simulation containing a $10^9 M_\odot$ chirp mass binary but beginning the sampling with a multi-resolution mesh which has $N_\text{side}=64$ at the location of the CW and decreases in resolution to $N_\text{side}=1$ following a multivariate normal, produced using a built-in \texttt{mhealpy} method and documented in its set of tutorials. We show in the bottom right panel of Figure \ref{fig:diagnostics} the trace plot of the number of pixels at each iteration of the algorithm for the two different initial meshes. As can be seen, the sampler moves efficiently across models until convergence. This shows that our sampler is robust to being initialized in poor regions of the space of models and that our inference is unaffected by starting with a convenient mesh.

\subsection{Information Criterion Analysis Results}
\begin{figure}
    \includegraphics[width=\columnwidth]{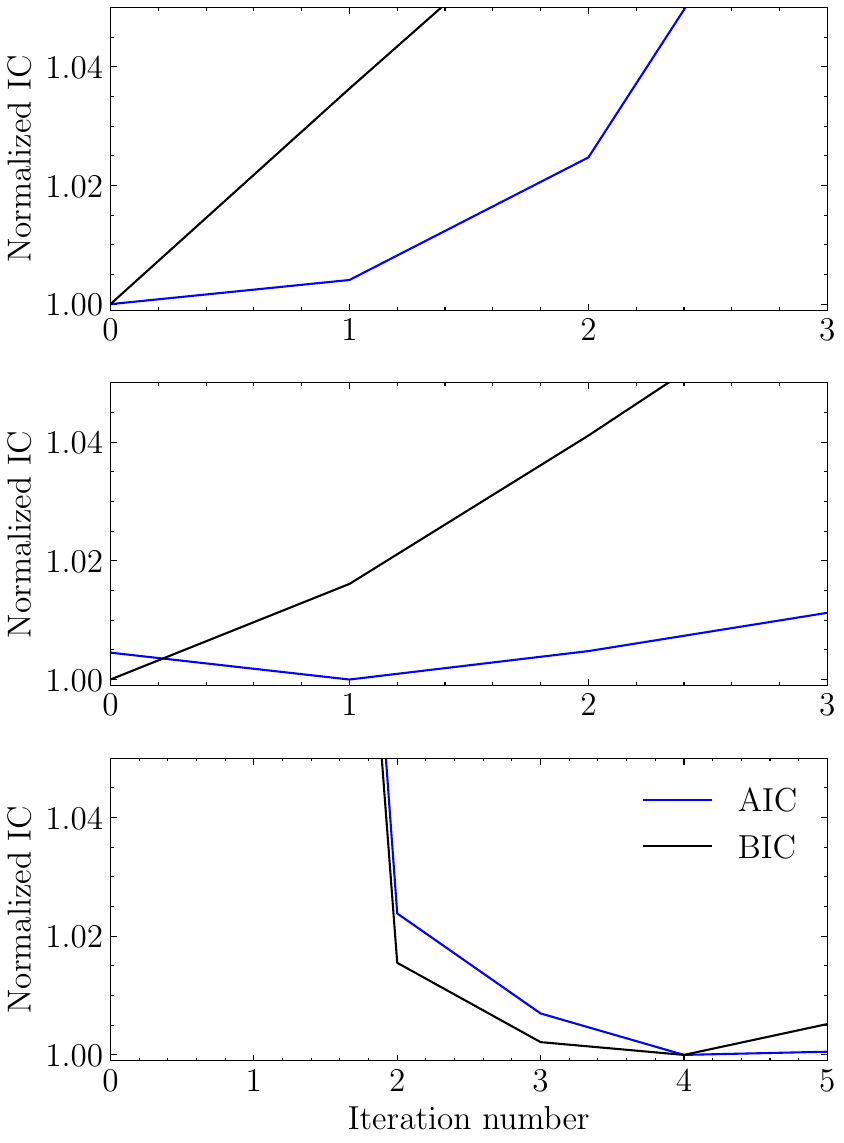}
    \caption{The AIC and BIC divided by their respective minimum value over several iterations of the frequentist method described in Section \ref{sec:bic}. The panels correspond to the realistic simulations having no CW, a $10^9 M_\odot$ chirp mass binary, and a $10^{10} M_\odot$ chirp mass binary, respectively, from top to bottom. The minimum AIC and BIC is attained after four iterations for the $10^{10} M_\odot$ chirp mass binary simulation, but the minimum AIC and BIC correspond to zero iterations for the simulation with no injected CW. For the simulation with a $10^9 M_\odot$ binary, the two information criteria disagree slightly, with the BIC favoring a base resolution HEALPix tessellation and the AIC favoring one iteration more. We show the corresponding preferred meshes and MLE angular power densities in Figure \ref{fig:bic_meshes}. The first two information criteria values for the bottom panel are not shown as we truncate the y-axis for legibility.}
    \label{fig:bic_minima}
\end{figure}

\begin{figure*}
    \includegraphics[width=\textwidth]{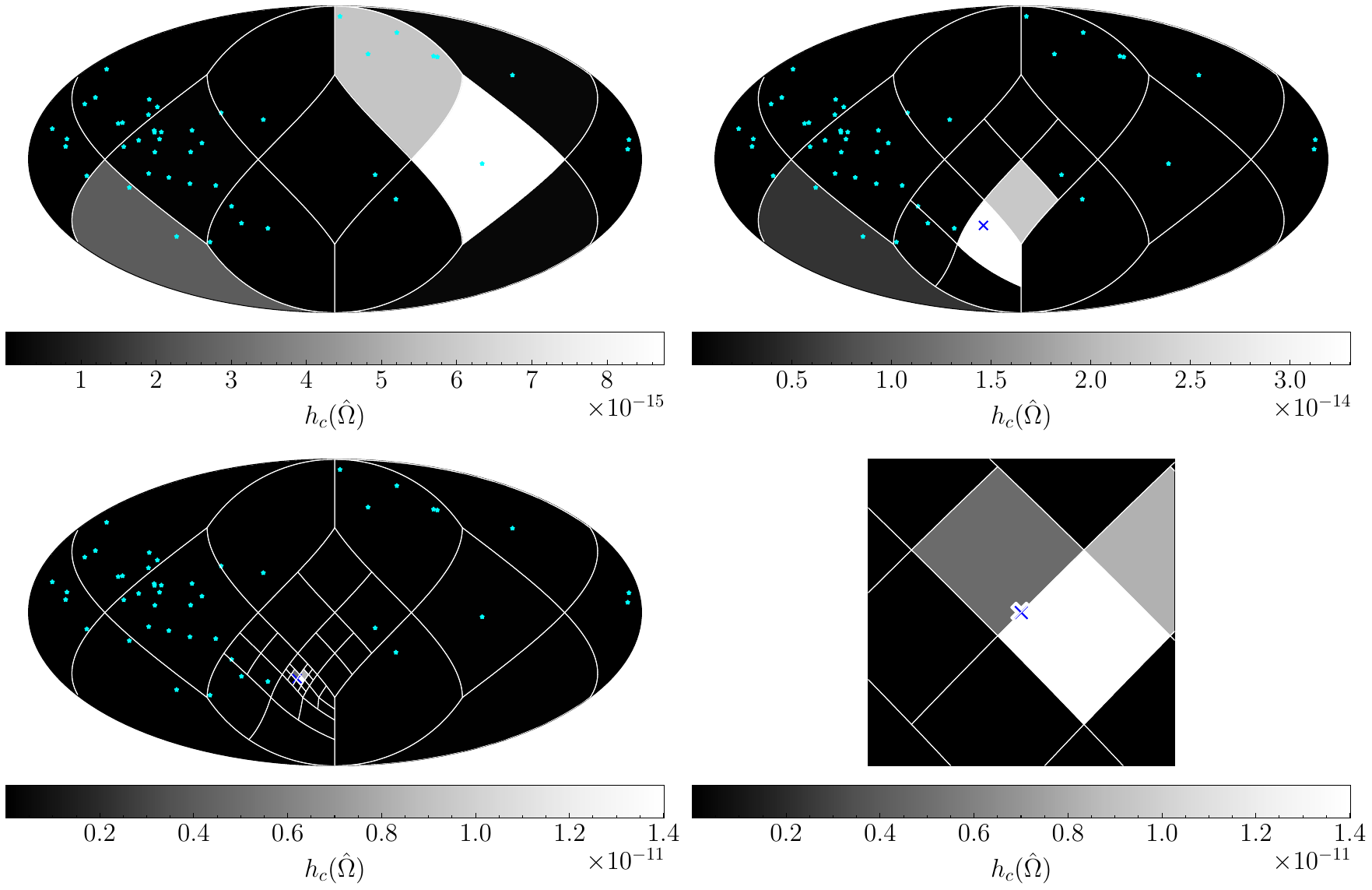}
    \caption{The preferred meshes and MLE angular power densities from the AIC results shown in Figure \ref{fig:bic_minima}. The top left Mollweide projection corresponds to the GWB-only simulation, the top right corresponds to the $10^9 M_\odot$ chirp mass binary injection, and the bottom row corresponds to the $10^{10} M_\odot$ chirp mass binary injection, with the left being an all-sky view and the right being a $10^{\circ}\times10^{\circ}$ Cartesian projection centered on the sky location of the injected CW.}
    \label{fig:bic_meshes}
\end{figure*}

\begin{figure}
    \includegraphics[width=\columnwidth]{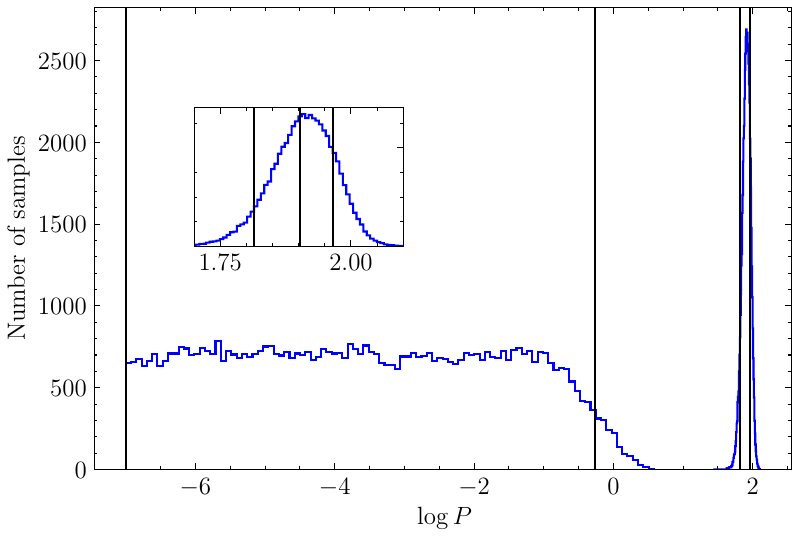}
    \caption{Comparison of frequentist uncertainties using profile likelihoods with 1d-marginalized posteriors from RJMCMC sampling for the realistic simulation containing an injected $10^9 M_\odot$ chirp mass binary. The blue histograms represent the marginalized posteriors, and the black vertical lines represent the frequentist $1-\sigma$ levels. The inset plot shows the posterior for the pixel containing the sky location of the injected CW, and the black vertical lines represent the $1-\sigma$ levels along with the MLE angular power density for that pixel. This shows validation of our frequentist uncertainties and also shows that these methods are able to capture asymmetries in the uncertainties. For legibility, we only show one upper-limit posterior.}
    \label{fig:uncertainties}
\end{figure}

To verify the rapid frequentist method proposed in Section \ref{sec:bic}, we use the method to analyze the realistic set of three simulations. We perform a few iterations for each simulation and show both the AIC and BIC as a function of iteration number in Figure \ref{fig:bic_minima}, although the values shown are normalized by dividing by the minimum value for each respective statistic. We present the preferred models and the MLE angular power densities in Figure \ref{fig:bic_meshes}. For one simulation, we also show the estimated $1-\sigma$ uncertainties on the angular power densities in Figure \ref{fig:uncertainties} overlaid on histograms of the inferred angular power densities from the corresponding RJMCMC analysis.

The preferred mesh is the same whether one uses the AIC or BIC for two of the simulations, but the preferred mesh is different between the two statistics for the simulation which contains a $10^9 M_\odot$ chirp mass binary. In this case, the mesh preferred using the AIC almost matches that of the RJMCMC analysis but contains an extra split pixel. This is because of the choice of threshold percentage of the integrated angular power density we use in this example; repeating the analysis with a lower threshold results in the same mesh being selected using the AIC as the mesh preferred from the RJMCMC analysis. In the case of the simulation with no injected CW, the two information criteria and the RJMCMC analysis indicated the same preferred model, the base resolution HEALPix tessellation. For the simulation containing a $10^{10} M_\odot$ chirp mass binary, the preferred model is the same whether one uses the AIC or BIC, but this model is different than the model preferred through an RJMCMC analysis as can be seen in Figure \ref{fig:realistic_results}. This is because the frequentist method we use increases the resolution at the locations of highest integrated angular power density and so cannot explore models which contain higher resolution pixels with low integrated angular power density unlike the RJMCMC sampler which is more flexible.

\section{Discussion and Conclusion} \label{sec:discussion}
Our set of simulation tests show that the sampler performs well; as shown in Figure \ref{fig:diagnostics}, it achieves good mixing both within models and between models, and it accurately recovers injected angular power densities. Additionally, the choice of initial mesh does not bias the inference; the sampler moves efficiently across model space towards the mode of the posterior as seen in the bottom right panel of Figure \ref{fig:diagnostics}.

The preferred mesh for the two simulations representative of isotropic GWBs -- one at the cross-correlation level with idealized noise and one at the TOA level with realistic noise levels -- is the base resolution HEALPix mesh. This is a natural consequence of Bayes' theorem, whereby models with fewer parameters are preferred over more complex models which do not fit the data substantially better. Using an RJMCMC sampler enables automatic application of this principle and therefore avoids overfitting (and underfitting). In contrast, if one were to use a counting argument and set $N_\text{side}=8$ (768 pixels) for the realistic simulations (which had 990 pulsar pairs) and $N_\text{side}=15$ (2700 pixels) for the idealistic simulations (which had 2775 pulsar pairs), this would result in overfitting the data in every sky location for most of the simulations and would even result in underfitting for the dataset with the $10^{10} M_\odot$ chirp mass binary injected. An additional, related benefit of this method is that the number of parameters in the sky map decreases by approximately one to three orders of magnitude, drastically reducing computational expense.

We also developed an alternative, significantly faster frequentist method based on information criteria.
While being much faster than an RJMCMC analysis, it loses the purely data-driven nature of the RJMCMC algorithm and introduces several arbitrary choices. In particular, one must choose a method for exploring models as it is computationally infeasible to compute the information criterion for every mesh since the model space is so numerous. We follow the method of \citet{Singer16} in that we start at a uniform resolution and iterate by splitting some user-specified threshold percentage of the integrated angular power density. In addition to being less data-driven, this method does not allow for straightforward model comparison between each pair of meshes; in contrast, the RJMCMC sampler automatically produces Bayes factors estimates for every pair of meshes explored.

Despite these flaws, we find the method works well in rapidly approximating the preferred mesh that the more robust sampling produces. This can be seen by comparing Figures \ref{fig:realistic_results} and \ref{fig:bic_meshes}. The maximum-likelihood point estimates for the angular power density are generally consistent with the posteriors, and the frequentist uncertainties on the angular power density are also in good agreement with the posteriors as can be seen in Figure \ref{fig:uncertainties}. The profile likelihood method we use for uncertainties is able to capture asymmetries in the uncertainties and allows it to be applied to upper-limit type pixels whereas a Fisher estimate would likely perform poorly in such cases (and cannot produce asymmetric uncertainties). As with Fisher estimates and Bayesian inference, the method is able to account for covariances between pixels since the values of other pixels are optimized as functions of the pixel under consideration.

Some of the results for the realistic simulations may seem unintuitive. In particular, the simulation with no CW injected results in strong overdensities of angular power in parts of the sky despite the injected GWB being isotropic. This result is consistent between the sampler and the frequentist method. To verify this is not unique to our new methods, we calculate a standard frequentist MLE with the square-root spherical harmonic basis \citep{Banagiri21,Payne20,Taylor20} as in, e.g., \citet{NG15anis,Gersbach25} with the same set of cross-correlations and covariance matrix, and we find the same overdensity in the recovered sky map, indicating this is intrinsic to the simulated data and is not an issue with our methods.

Another result that may seem unintuitive is that, for the realistic simulation with a $10^{10} M_\odot$ chirp mass binary injected, the preferred mesh selected through our RJMCMC sampler may appear to be a worse model than the preferred mesh selected through the AIC and BIC since the pixel of maximum angular power density does not contain the injected CW and because there are regions of the sky with low angular power density but higher resolution. As a quick test, we perform numerical optimization of the likelihood for each of the two preferred meshes and find that the mesh selected by RJMCMC corresponds to a maximum log-likelihood greater than that of the AIC and BIC mesh by $\sim100$. These counterintuitive effects are likely due to realization-dependent noise fluctuations in the process of simulating the TOAs for each dataset.

Finally, we note that our sampler and methods do not depend fundamentally on the frequentist cross-correlation likelihood in Equation \ref{eq:likelihood}, so one could use a Bayesian likelihood at the level of the TOAs instead of the likelihood we use in this work. However, the sampler does require \texttt{JAX}-based likelihoods, e.g. for automatic differentiation, so this likelihood could be constructed with \texttt{Discovery} \citep{discovery}.

\begin{acknowledgments}
T.T.M. thanks Rutger van Haasteren for sandboxing \texttt{libstempo}. T.T.M and N.S.P. acknowledge support from startup funds from Texas Tech University. The authors acknowledge the High Performance Computing Center (HPCC) at Texas Tech University for providing computational resources that have contributed to the research results reported within this paper. Some of the results in this paper have been derived using the \texttt{healpy} and \texttt{HEALPix} packages.
\end{acknowledgments}

\begin{contribution}
T.T.M. conceived of the project, implemented the sampler, performed the analyses, produced the plots, and wrote the manuscript. N.S.P. provided guidance throughout.
\end{contribution}

\software{
          \texttt{Defiant} \citep{pfos},
          \texttt{ENTERPRISE} \citep{enterprise},
          \texttt{enterprise\_extensions} \citep{enterprise_extensions},
          \texttt{FastShermanMorrison} \citep{fastshermanmorrison},
          \texttt{healpy} \citep{Zonca2019},
          \texttt{JAX} \citep{jax2018github},
          \texttt{JAXopt} \citep{jaxopt_implicit_diff},
          \texttt{la\_forge} \citep{laforge},
          \texttt{libstempo} (https://github.com/vallis/libstempo),
          \texttt{MAPS} \citep{Pol22},
          \texttt{Matplotlib} \citep{Hunter:2007},
          \texttt{mhealpy} \citep{mhealpySoftware},
          \texttt{Numba} \citep{Numba},
          \texttt{NumPy} \citep{harris2020array},
          \texttt{Optimistix} \citep{optimistix24},
          \texttt{PTMCMCSampler} \citep{ptmcmcsampler},
          \texttt{Resolve} (https://github.com/TTMoursy/Resolve),
          \texttt{SciPy} \citep{SciPy}
}

\bibliography{references}{}
\bibliographystyle{aasjournalv7}

\end{document}